\documentclass{ws-procs9x6}

\begin{document}

\title{Kaon Photoproduction in the Feynman and Regge Theories}

\author{T. Mart}

\address{Departemen Fisika, FMIPA, Universitas Indonesia, Depok 16424, Indonesia}

\author{C. Bennhold}

\address{Center for Nuclear Studies, Department of Physics, The George 
         Washington University, Washington, D.C. 20052, USA}  

\maketitle

\abstracts{Kaon photoproduction on the nucleon has been investigated in the 
isobar (Feynman) and Regge frameworks. Two possible combinations of the theories
are discussed. Results are compared with present experimental data.
}

\section{Introduction}
A significant progress has been achieved in the experimental side of 
the strangeness photoproduction. A wealth of new high-statistics data 
on elementary kaon photoproduction has recently become 
available in all three isospin channels.\cite{Glander:2003jw,Goers:1999sw,McNabb:2003nf}
Along with some new advancements in the theoretical side this
has made the field of kaon electromagnetic production to be of
considerable interest. Nevertheless, in spite of substantial efforts
spent for almost 40 years, a comprehensive and consistent description
of the underlying reaction mechanism is far from available. This
might be due to the presence of the strangeness, which explicitly appears
in the final state of the process. The presence of this additional degree
of freedom leads to a more complex theoretical effort in explaining 
experimental data. One of the most tantalizing problem is the high
energy reaction threshold. Thus, even at threshold, a copious number of
baryon resonances might already contribute to the process.

On the other hand, an extension of phenomenological models to higher 
energies becomes an urgent 
task if, for instance, we consider the calculation of the GDH sum rule.
At Jefferson Lab, kaon electroproduction experiment has been 
performed with total c.m. energy $W =3$ GeV. A proposal for upgrading 
the accelerator to reach 12 GeV has been also discussed.\cite{jlab1998}
To this end, no isobar model has been proposed to investigate physics 
of the process at this energy. 

It is the purpose of this paper to discuss such an extension by using the 
available achievements we currently have accomplished, i.e., the isobar 
(Feynman) and Regge models. Here we try to combine both models in order to explain
kaon photoproduction data from threshold up to $E_\gamma^{\rm lab}=16$ GeV.

\section{Formalism}
\subsection{Feynman Amplitude}
For the sake of simplicity we will only consider $K^+\Lambda$ photoproduction
to elucidate the main features of the present investigation.
With the convention of four-momenta
\begin{eqnarray}
\gamma (p_\gamma) + p (p_p) \longrightarrow K^+(q) + \Lambda(p_\Lambda) ~,
\end{eqnarray}
the transition matrix for both isobar and Regge models for kaon
photoproduction can be written in the form of
\begin{eqnarray}
M_{\mathrm fi} &=& {\bar u}({\mbox{\boldmath ${p}$}}_\Lambda,s_\Lambda) 
  \sum_{i=1}^{4} A_i~M_i ~u({\mbox{\boldmath ${p}$}}_p,s_p) ~,
\label{eq:mfi}
\end{eqnarray}
where the gauge and Lorentz invariant matrices $M_i$ are given 
by\cite{Lee:1999kd,Mart:2003yb}
\begin{eqnarray}
 M_1 &=& \gamma_5 ~\epsilon\;\!\!\!\!/ p_\gamma\!\!\!\!\!\!/ ~~,\\
 M_2 &=& 2\gamma_5 (p_K\cdot\epsilon ~p_p\cdot p_\gamma - p_K\cdot p_\gamma~
         p_p\cdot\epsilon) ~,\\
 M_3 &=& \gamma_5 (p_K\cdot p_\gamma~ \epsilon\;\!\!\!\!/ - p_K\cdot \epsilon~
         p_\gamma\!\!\!\!\!\!/ ~ ) ~, \\
 M_4 &=& i\varepsilon_{\mu\nu\rho\sigma}\gamma^\mu p_K^\nu \epsilon^\rho
         p_\gamma^\sigma ~,
\end{eqnarray}
The functions $A_i$ are obtained from 
the appropriate Feynman diagrams and for the isobar model we use the
same resonance configuration as in Ref.\cite{Mart:2000ed}, where hadronic form 
factors are included by utilizing the method of Ref.\cite{Haberzettl:1998eq}
to restore gauge invariance.

\subsection{Contact Terms}
Contact terms are usually included to restore gauge invariance of the
reaction amplitudes\cite{Haberzettl:1998eq}. However, there is no 
restriction for using these
terms completely. The most general interaction Lagrangian for the contact terms in 
meson photoproduction can be written as
\begin{eqnarray}
  \label{eq:lagrangian}
\mathcal{L}_{\gamma\phi BB} &=& \bar{\psi}\partial_{\mu}\phi
F_{\alpha\beta}\left\{\frac{\beta_1}{m}[ \gamma^\mu
\gamma^\alpha \gamma^\beta - \gamma^\mu g^{\mu\nu}] +
\frac{\beta_2}{m} [g^{\mu\alpha}\gamma^\beta -
g^{\mu\beta}\gamma^\alpha]  + \right.\nonumber \\ &&\frac{\beta_3}{m
M}[ P^\mu (\gamma^\alpha \gamma^\beta - g^{\alpha\beta})] +
\frac{\beta_4}{mM^2} [g^{\alpha\beta} P^\alpha
-g^{\mu\beta} P^\alpha] + \nonumber \\ &&\left.\frac{\beta_5}{mM}
[P^\mu (P^\alpha \gamma^\beta - \gamma^\alpha P^\beta]
\right\}\gamma_5\psi ~,
\end{eqnarray}
where $\phi~ (\psi)$ is the meson (baryon) field, $m~(M)$ refers to the meson 
(baryon) mass, $\beta_i$ is the corresponding coupling constant,
$F_{\alpha\beta}=\partial_\alpha A_\beta -\partial_\beta A_\alpha$, 
and $P=p_B+p_{B'}$. Using Eq.\,(\ref{eq:mfi}) we can decompose the
corresponding amplitudes to obtain 
\begin{eqnarray}
  \label{eq:contact_terms}
  A_1 &=& \frac{e\beta_3}{M^4}\left( s-u+m_p^2-m_\Lambda^2\right) +
  \frac{ie\beta_5}{M^5}\left( s-u+m_p^2-m_\Lambda^2\right)\left(m_p+m_\Lambda\right) 
  \, ,~~~\\
  A_2 &=& \frac{2e\beta_4}{M^4} ~,\\
  A_3 &=& \frac{2ie}{M^3}\left(\beta_1+\beta_2\right) ~,\\
  A_4 &=& -\frac{2ie\beta_1}{M^3}+\frac{ie\beta_5}{M^5}\left(s-u+m_p^2-
    m_\Lambda^2\right) ~,
\end{eqnarray}
where $M=1$ GeV is taken in order to make $\beta_i$ dimensionless.
These functions belong to the isobar model.

\subsection{Regge Amplitudes}
The procedure to obtain the photoproduction amplitude for the Regge model 
is adopted from Ref.\cite{guidal97}, i.e., by 
replacing the Feynman propagator with the Regge propagator
\begin{eqnarray}
  \label{eq:regge}
  P_{\rm Regge} &=& \frac{s^{\alpha_{K^i}(t)-1}}{\sin [\pi\alpha_{K^i}(t)]}
                    ~ e^{-i\pi\alpha_{K^i}(t)} ~ 
                    \frac{\pi\alpha_{K^i}'}{\Gamma [\pi\alpha_{K^i}(t)]} ~,
\end{eqnarray}
where $K^i$ refers to $K$ and $K^*$, and 
$\alpha_{K^i} (t) = \alpha_0 + \alpha '\, t$ denotes the 
trajectory\cite{guidal97}. The extracted functions $A_i$ are 
given in Ref.\cite{Mart:2003yb}.

\subsection{Combining Feynman and Regge Amplitudes}
In order to explain kaon photoproduction from threshold up to 
$E_\gamma^{\rm lab}=16$ GeV ($W\approx 5$ GeV) we can combine Feynman 
and Regge models by mixing $A_i$ obtained from Eq.\,(\ref{eq:mfi}).
Following Ref.\,\cite{Barbour:1978cx} the procedure reads 
\begin{eqnarray}
  \label{eq:NPA}
  A_i &=& \frac{1}{s_1-s_2} \, \left\{(s-s_2)\ A_i^{\rm iso} +
          (s_1-s)\ A_i^{\rm Reg}\, \right\} ,  ~~~~
         i=1,...,4 
\end{eqnarray}
where $\sqrt{s_1}$ and $\sqrt{s_2}$ define the transition region 
in which both models are mixed. We note that this procedure has 
been successfully applied to a multipole analysis of single-pion 
photoproduction between threshold and $E_\gamma^{\rm lab}=16$ GeV. 
In our calculation we use $\sqrt{s_1}=W_{\rm thr}$ and 
$\sqrt{s_2}=2.5$ GeV. Therefore, all low and medium energy data
fall in this transition region. The result obtained by using this 
method will be indicated as ``mixed 1''.

Alternatively, one can use
\begin{eqnarray}
  \label{eq:exp}
  A_i &=& e^{-(s-s_{\rm thr})/a}\ A_i^{\rm iso} +
          \left\{1-e^{-(s-s_{\rm thr})/a}\right\} A_i^{\rm Reg} ,  ~~~~
         i=1,...,4 ~~~~
\end{eqnarray}
where thr refers to threshold and $a$ is a fitted constant.
Clearly, compared to the previous method, 
the advantage of using Eq.\,(\ref{eq:exp}) is that it ensures a smooth 
transition of amplitudes from the isobar to the Regge regimes. In the
subsequent discussion we will call this method as ``mixed 2''.

\section{Results and Discussion}
\begin{figure}[t]
\centerline{\epsfxsize=3.3in\epsfbox{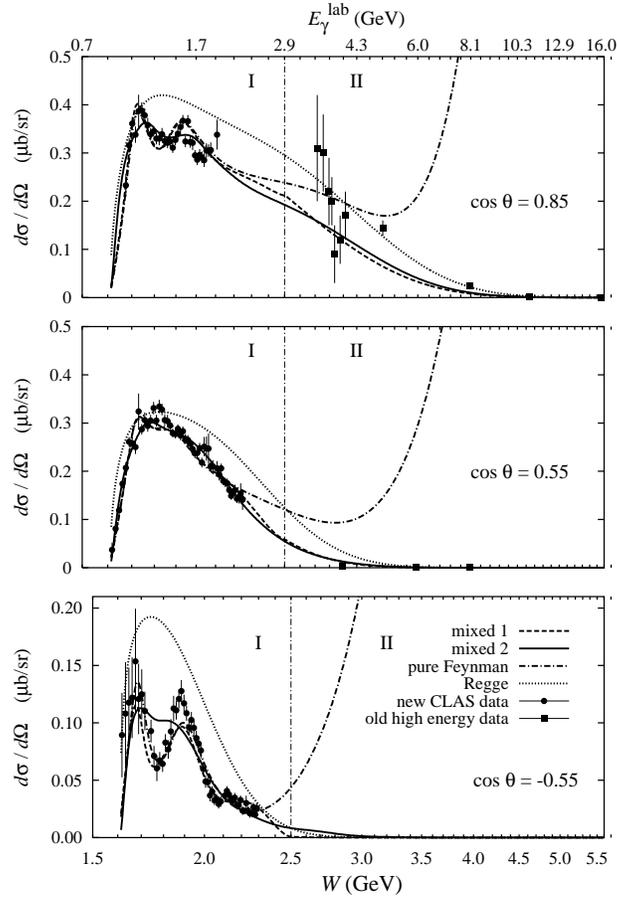}}   
\caption{Differential cross sections for the $\gamma +p\to K^+ +\Lambda$
  channel obtained by using different models. 
  Data are taken from Ref.\protect\cite{McNabb:2003nf} (solid circles) and 
  Ref.\protect\cite{old_data} (solid squares).
  \label{fig:dcs}}
\end{figure}

The results of our calculation are displayed in Fig.\,\ref{fig:dcs}, where
we compare differential cross sections, for three different kaon angles, 
obtained from the isobar, Regge, mixed 1, and mixed 2 models.

\begin{figure}[t]
\centerline{\epsfxsize=4.9in\epsfbox{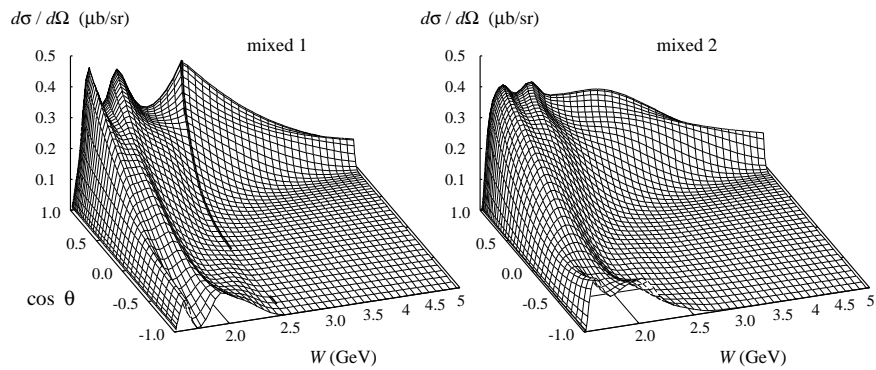}}   
\caption{Differential cross section as functions of $\cos\theta$
  and $W$ for the two mixed models. Transition from the isobar to
  the Regge regimes is market by a solid line at $W=2.5$ GeV in
  the case of ``mixed 1'' model.
  Notation is as in Fig.\,\ref{fig:dcs}.
  \label{fig:mixed}}
\end{figure}

As has been expected, the cross section obtained from the Regge model works
nicely only at high energies, a domain where the isobar one appears to be divergent,
while a contrary situation happens at low energies. By combining the two models
using Eq.\,(\ref{eq:NPA}) good descriptions of low energy and high energy
data can be achieved. However, the shortcoming of this method appears
obviously at the ``transition'' point, i.e. $W=2.5$ GeV, where the fit
switches from a mixed to a pure Regge model. The discontinuity
at this point leads to the major deficiency of this method, especially at 
forward angles as will be shown later in Fig.\,\ref{fig:mixed}.

On the other hand, Eq.\,(\ref{eq:exp}) yields a smooth transition from
the isobar to the Regge model. The only problem of such mixing method is
observed at low energy and backward angles, where the model is not
able to reproduce the two observed peaks. The problem is originated from the
relatively large contribution of Regge amplitudes. As can be seen from
Fig.\,\ref{fig:dcs}, Regge model tends to flatten and enhance the cross
section. The best fit obtained by using this model yields a mixing 
parameter $a=1.73$ GeV$^2$ with $\chi^2/N=2.61$.
It can be easily calculated from Eqs.\,(\ref{eq:NPA}) and (\ref{eq:exp}) that
a same contribution from the Regge model would be obtained in ``mixed 2'' if
we used $a=2.74$ GeV$^2$. Fixing this value in the ``mixed 2'' model leads to 
a larger value of $\chi^2/N$, i.e., 2.96.

\begin{figure}[t]
\centerline{\epsfxsize=3.5in\epsfbox{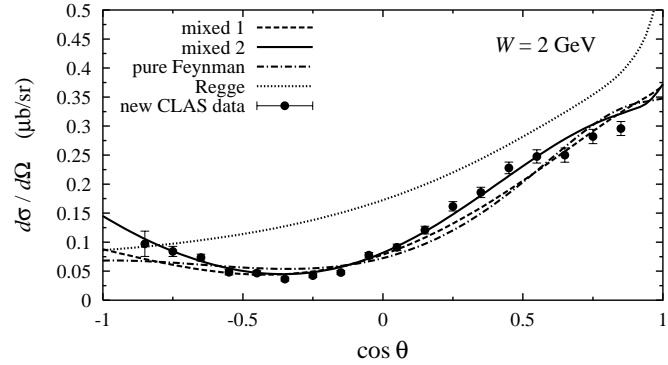}}   
\caption{Angular distribution of the differential cross section 
  obtained by using different models. Notation is as in
  Fig.\,\ref{fig:dcs}.
  \label{fig:dcs_th}}
\end{figure}

Figure \ref{fig:dcs_th} shows the angular distribution of the differential
cross section. Clearly, only the ``mixed 2'' model can give the best
explanation of experimental data, especially at the very backward angle,
where a small increment of the cross section is detected at this region.
This result is interesting since the shape of angular distributions of the observable
is usually determined by the $t$-channel contributions.

\begin{figure}[t]
\centerline{\epsfxsize=3.5in\epsfbox{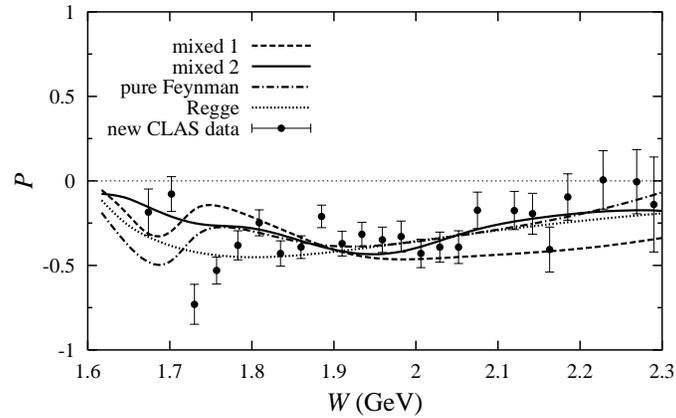}}   
\caption{Recoil polarizations obtained by using different models. 
  Notation is as in Fig.\,\ref{fig:dcs}.
  \label{fig:polar}}
\end{figure}

A relatively uniform result is obtained for the recoil polarization.
We observe that no model can reproduce this polarization observable accurately, 
especially the interesting structure at $W\approx 1.75$ GeV. This could be a sign that 
another resonance plays an important role at this energy. Future calculation should
address this question. At this moment, in view of 
the obtained $\chi^2$ and the error bars of the polarization data, these 
results are still understandable. 

\begin{figure}[t]
\centerline{\epsfxsize=3.5in\epsfbox{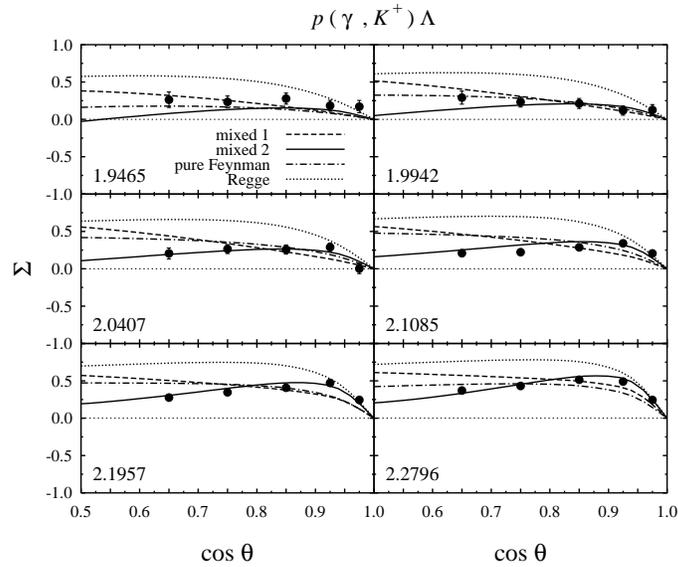}}   
\caption{Photon asymmetry observables obtained by using different models. 
  Notation for the curves is as in Fig.\,\ref{fig:dcs}. Data are
  taken from Ref.\protect\cite{Zegers:2003ux}.
  \label{fig:polpho}}
\end{figure}

The photon asymmetry data again show that the ``mixed 2'' model is 
superior to others. As shown in Fig.\,\ref{fig:polpho} a slight decrement 
of the observable is reproducible only by the ``mixed 2'' model,
whereas other models tend to increase the polarization as $\cos\theta$
decreases.

Finally, in Fig.\,\ref{fig:mixed} we show a three-dimensional view
of the predicted differential cross sections. It is obvious that a
discontinuity exists along the line of $W=2.5$ GeV and reaches a maximum at
the forward angle. In contrast to this a smooth cross section
surface is displayed by the ``mixed 2'' model. A strong Regge
effect appears in the resonance region; the cross section tends to become more flat,
thus reducing the effect of resonances at the two peaks.
This effect is of course not demanded by the present experimental
data. 

In conclusion we have investigated kaon photoproduction from threshold 
up to $E_\gamma^{\rm lab}=16$ GeV by combining an isobar and a Regge models
in two different methods.
We have shown that both methods presented here have different shortcomings.
Future investigations should try to remedy this problem by using, e.g.,
additional nucleon resonances or different mixing recipes.

\section*{Acknowledgments}
The work of TM has been supported in part by the QUE project.

\end{document}